\documentclass[fp,twocolumn]{jpsj3}
\usepackage[pdftex, colorlinks=true, linkcolor=blue, citecolor=blue, urlcolor=blue]{hyperref}
\usepackage{graphicx}
\usepackage{here}
\usepackage{color}
\usepackage{txfonts}
\usepackage{bm}
\usepackage[monochrome]{xcolor}

\title{Simulated Annealing for Quadratic and Higher-Order Unconstrained Integer Optimization}

\author{Kohei Suzuki}
\inst{Jij Inc., 3-3-6 Shibaura, Minato-ku, Tokyo 108-0023, Japan} 

\abst{
Simulated annealing (SA) is a key algorithm for solving combinatorial optimization problems, which model numerous real-world systems.
While SA is commonly used to solve quadratic unconstrained binary optimization (QUBO) problems, many practical problems are more naturally formulated using integer variables.
We therefore study quadratic and higher-order unconstrained integer optimization (QUIO and HUIO) problems, which generalize QUBO by allowing integer-valued variables and higher-order interactions.
Conventional approaches often convert these problems into QUBO formulations through binary encoding and the reduction of higher-order terms.
Such conversions, however, greatly increase the number of variables and interactions, resulting in long computation times and, for large-scale problems, even making the conversion itself a dominant computational bottleneck.
To overcome this limitation, we propose an efficient framework that directly applies SA to QUIO and HUIO problems without converting them into QUBO.
Within this framework, we introduce an optimal-transition Metropolis method, designed to improve efficiency when the variable range is wide, and evaluate its performance alongside the conventional Metropolis and heat bath methods.
Numerical experiments demonstrate that the proposed direct approach achieves higher efficiency and better solution quality than the conventional QUBO-based formulation and reveal the practical advantages of the optimal-transition Metropolis method.
The algorithm developed in this study is available as part of the open-source library OpenJij, which provides a Python front end with a C++ backend.
}

\begin{document}
\maketitle

\section{Introduction}
Combinatorial optimization problems play a central role in a wide range of real-world applications, including finance, logistics, drug discovery, and scheduling.
Many of these problems are NP-complete or NP-hard, and obtaining exact solutions in polynomial time is generally intractable.\cite{fphy.2014.00005}
Therefore, efficient approximation algorithms, often referred to as metaheuristics, are indispensable.
Among them, simulated annealing (SA)\cite{SA, SA2}, a highly versatile metaheuristic inspired by the analogy with the physical annealing process, is one of the most powerful and widely used approaches due to its ability to escape from local optima and search for the global optimum.

When applying SA to combinatorial optimization problems, it is common to reformulate them as quadratic unconstrained binary optimization (QUBO) problems.\cite{fphy.2014.00005, Kochenberger2014, Glover2022}
A QUBO problem is defined as minimizing a quadratic polynomial whose variables take binary values $\{0, 1\}$.
However, in many real-world cases, modeling the problem with integer variables is more natural than with binary variables.
In such cases, it is natural to formulate the problem as a quadratic unconstrained integer optimization (QUIO) problem.
Furthermore, these problems often contain higher-order polynomial terms or constraints, which can be represented by extending QUIO to the higher-order unconstrained integer optimization (HUIO) formulation.

In previous studies, QUIO and HUIO problems have often been solved by converting them into QUBO formulations.
However, this conversion inherently introduces two major computational bottlenecks.
First, binary encoding is required to transform each integer variable into multiple binary variables.\cite{Karimi2019, 9435359}
Second, when the objective function contains higher-order terms beyond quadratic, higher-order term reduction must be carried out to express them in quadratic form.\cite{BOROS2002155, Anthony2017}
These conversion procedures dramatically increase both the number of variables and the number of interactions, thereby making the resulting QUBO model much denser and its search space more complex.
For large-scale problems, the conversion process itself can become a dominant computational overhead, and in some cases it does not finish within a practical amount of time, which makes it a serious limitation in practice.
Therefore, it is desirable to solve such problems without converting them into QUBO.
Although SA can in principle be applied directly to QUIO and HUIO formulations, actual applications of this approach remain scarce. 
Existing studies address only limited classes of QUIO problems\cite{CARDOSO19971349, Abramson1999, integer_sa_1}, and research on the general HUIO case appears to be extremely scarce.\cite{TIAN1995629}

In this study, we propose an efficient framework that directly applies SA to general QUIO and HUIO formulations without converting them into QUBO, and we describe in detail how SA can be implemented efficiently for such problems.
In typical SA implementations for QUBO problems, a single-spin flip update, where one binary variable is flipped at a time, is commonly used.
Similarly, in our framework, a single-variable update is employed, where only one integer variable is changed at a time.
Under this scheme, the energy difference required for variable updates, which is often a computational bottleneck in SA, can be efficiently computed even for QUIO and HUIO formulations.
As for transition probabilities, the Metropolis method\cite{METROPOLIS, METROPOLIS_HASTING} and the heat bath method\cite{10.1071/PH650119, 4767596} are widely used; however, when the range of integer variables are large, both approaches encounter difficulties.
To overcome this issue, we introduce the optimal-transition Metropolis method, which enables efficient updates under wide variable range.
Using these transition schemes, we apply SA directly to QUIO and HUIO and evaluate their performance, including comparisons with the conventional QUBO-based approach.

This paper is organized as follows.
Section \ref{sect-model} presents the formulations of QUIO and HUIO.
Section \ref{sect-sa} describes the proposed SA algorithm in detail, with particular emphasis on the computation of energy differences and transition probabilities.
Section \ref{sect-results} reports the numerical results, and Section \ref{sect-summary} concludes the paper with a discussion of future perspectives.

\section{Model}\label{sect-model}
In this section, we introduce the polynomial unconstrained integer optimization, namely QUIO and HUIO.
We begin with the quadratic unconstrained integer optimization.
QUIO is defined as the problem of minimizing a quadratic function composed of integer variables:
\begin{align}
    E_{p=2}(\bm{z})
    = \sum_{i_1} J_{i_1} z_{i_1} + \sum_{i_1,i_2} J_{i_1,i_2} z_{i_1} z_{i_2}.
\label{def_QUIO}
\end{align}
Here, $\bm{z}=(z_1, z_2, \ldots, z_N)$ is the set of integer variables and $N$ is the number of variables.
Each variable satisfies $l_i \le z_i \le u_i$, and $l_i$ and $u_i$ denote its lower and upper bounds, respectively.
Note that terms such as $J_{i_1,i_2} z_{i_1} z_{i_2}$ and $J_{i_2,i_1} z_{i_2} z_{i_1}$ have already been merged into a single term.
Thus, without loss of generality, the index pair $(i_1,i_2)$ in $J_{i_1,i_2}$ can be regarded as an unordered pair,
so that $J_{i_1,i_2}$ and $J_{i_2,i_1}$ denote the same coefficient.

When $l_i = 0$ and $u_i = 1$ for all $i$, the variables take binary values $z_i \in \{0,1\}$, and the problem reduces to the QUBO problem, which is widely known in combinatorial optimization.
QUIO can thus be regarded as a natural generalization of QUBO from binary to integer variables, and, as in the QUBO case, many optimization problems can be formulated as QUIO.

Next, we introduce the higher-order unconstrained integer optimization, HUIO,
which is defined as the problem of minimizing a polynomial that includes terms of order three or higher:
\begin{align}
    E_p(\bm{z})
    =&\sum_{i_1}J_{i_1} z_{i_1} + \sum_{i_1,i_2}J_{i_1,i_2} z_{i_1} z_{i_2} + \sum_{i_1,i_2,i_3}J_{i_1,i_2,i_3} z_{i_1} z_{i_2} z_{i_3}+\nonumber\\
    &\cdots+\sum_{i_1,i_2,\ldots,i_p}J_{i_1,i_2,\ldots,i_p} z_{i_1} z_{i_2}\cdots z_{i_p}.
    \label{def_HUIO}
\end{align}
Here, $p$ denotes the polynomial order, and the case $p=2$ corresponds to QUIO.
As in the QUIO case, terms with the same index multiset have already been merged into a single term.
Thus, without loss of generality, the index tuple $(i_1,i_2,\ldots)$ can be regarded as unordered,
so that all its permutations denote the same coefficient $J_{i_1,i_2,\ldots}$.

In the case of binary variables, HUIO reduces to the higher-order unconstrained binary optimization (HUBO).\cite{IJMHEUR.2011.041196, IJMHEUR.2011.044356}
HUBO formulations also arise naturally in real-world optimization models, and there are studies in which SA has been applied directly to such cases.\cite{PhysRevA.109.032416, Wang2025, ikeuchi2025}
HUIO can be viewed as a natural extension of QUIO or HUBO, and higher-order terms of this form naturally arise in practical optimization problems, for example when handling quadratic constraints involving integer variables using penalty or augmented Lagrangian methods.\cite{Bertsekas1982}

In the context of mathematical optimization, $E_p(\bm{z})$ is referred to as the objective function, and its minimum value is called the optimal value, while any configuration $\bm{z}$ that gives this value is called an optimal solution. 
In general, multiple optimal solutions may exist that yield the same optimal value. 
From the viewpoint of statistical physics, $E_p(\bm{z})$ corresponds to the Hamiltonian of a classical spin system, where the variables $z_i$ can be regarded as generalized classical spin variables that take arbitrary integer values. 
In this interpretation, QUIO and HUIO represent the problem of finding the ground state of the spin system and its corresponding energy, and the existence of multiple optimal solutions corresponds to degeneracy of the ground state. 
In this paper, we use the terminology from optimization and physics appropriately depending on the context.

\section{Simulated Annealing}\label{sect-sa}
In this section, we present the SA framework for general QUIO and HUIO, and describe several techniques to make it efficient.
Section \ref{subsect-sa} outlines the SA procedure itself, including initialization, cooling schedule, and update rules.
Section \ref{subsect-energy-diff} then explains a method for efficiently evaluating the energy difference when a variable changes.
Section \ref{subsect-transi-prob} discusses different choices of transition probabilities, ranging from the standard Metropolis and heat bath methods to the proposed optimal-transition Metropolis method.
Finally, Section \ref{subsect-init-temp} describes how to determine appropriate initial and final temperatures for SA, providing a practical scheme for setting them based on the properties of the objective function.

\subsection{Algorithm}\label{subsect-sa}
This subsection provides an overview of the SA algorithm.
The procedure begins with the preparation of an initial state $\bm{z}=\bm{z}_{\text{init}}$, initialized at random.
The initial temperature is then set to $T=T_{\max}$, which is chosen to be sufficiently high relative to the energy scale of $E_p(\bm{z})$.
The following steps are repeated at a fixed temperature $T$:
\begin{enumerate}
    \item Select a variable $z_k$ and compute the energy difference $\Delta E_{p,k}$ associated with the variable change $z_k \to z_k + \Delta z_k$.
    \item Evaluate the transition probability $P(\Delta E_{p,k},T)$ and accept the change $z_k \to z_k + \Delta z_k$ according to this probability.
\end{enumerate}
Here, we call one pass over all variables, visiting $k=1,\ldots,N$ and applying steps (1)–(2) to each variable once, a sweep.
After each sweep, the temperature $T$ is decreased.
There exist several ways to update the temperature.\cite{SZU1987157, Ingber1996, YaghoutNourani_1998, S1052623497329683, Mahdi2017}
For example, a logarithmic cooling schedule provides a theoretical guarantee that SA converges to the global optimum\cite{4767596}, although the cooling is so slow that it is impractical for real-world computations.
In practice, geometric cooling is widely used due to its simplicity and good empirical performance.
In this study, we employ the following geometric cooling schedule:
\begin{align}
    T_i = T_{\max}\left(\frac{T_{\min}}{T_{\max}}\right)^{\frac{i-1}{N_{\text{sweep}}-1}},\qquad i=1,2,\ldots,N_{\text{sweep}} .
\end{align}
Here, $T_{\min}$ is the final temperature of SA, $T_i$ is the temperature at the $i$-th sweep, and $N_{\text{sweep}}$ is the total number of sweeps.

Although the variable to be updated can be chosen at random, sequential selection is often more effective in practice; accordingly, we visited variables in index order during each sweep.
The calculation of the energy difference $\Delta E_{p,k}$, the definition of the transition probability $P(\Delta E_{p,k},T)$, and the procedure for determining $T_{\max}$ and $T_{\min}$ are described in the following subsections.

\subsection{Energy difference}\label{subsect-energy-diff}
To perform SA efficiently, the energy difference when a variable changes must be calculated efficiently. 
A common method for QUBO problems is to precompute the energy difference for the flip of each variable and update these values when a flip is accepted.
In this approach, the energy difference can be obtained in $O(1)$ time, which is particularly advantageous when variable updates are infrequent. 
In fact, during the low-temperature stage of SA, the system is already close to its ground state and variable updates rarely occur. 
As a result, this method improves the computational efficiency.

However, this method is only applicable to binary variables, because they can transition to only one alternative state.
For QUIO and HUIO, an integer variable may transition to one of multiple possible states, and thus this strategy cannot be applied directly.
Instead of precomputing the energy differences themselves, we store the corresponding coefficients.

Let us first consider the case of QUIO [Eq. (\ref{def_QUIO})].
Suppose that a variable $z_k$ is updated to $z_k + \Delta z_k$.
The corresponding energy difference can then be expressed as quadratic form with respect to $\Delta z_k$:
\begin{align}
    \Delta E_{p=2,k}(\bm{z})
    &=E_{p=2}(\bm{z} + \Delta \bm{z}_k) - E_{p=2}(\bm{z})\nonumber\\
    &=c^{(1)}_k(\bm{z})\Delta z_k + c^{(2)}_k(\bm{z})(\Delta z_k)^2.
    \label{energy_diff_quad}
\end{align}
Here, $\Delta\bm{z}_k=(0,\ldots,\Delta z_k,\ldots,0)$.
The coefficients $c^{(1)}_k(\bm{z})$ and $c^{(2)}_k(\bm{z})$ can be calculated explicitly as
\begin{align}
    c^{(1)}_k(\bm{z})&=2J_{k,k}z_k + J_{k} + \sum_{\substack{i_1 \\ i_1 \neq k}}J_{i_1,k}z_{i_1}, \\
    c^{(2)}_k(\bm{z})&=J_{k, k}.
\end{align}
The important point is that both $c^{(1)}_k(\bm{z})$ and $c^{(2)}_k(\bm{z})$ are independent of the change $\Delta z_k$.
Therefore, once $c^{(1)}_k(\bm{z})$ and $c^{(2)}_k(\bm{z})$ are precomputed, the energy difference associated with a change of $\Delta z_k$ can be obtained in $O(1)$ time using Eq. (\ref{energy_diff_quad}).
Here, $c^{(2)}_k(\bm{z})$ is a constant independent of any variable $z_i$, and thus only $c^{(1)}_k(\bm{z})$ needs to be updated when a variable update is accepted.
Suppose that a variable $z_s$ is updated to $z_s + \Delta z_s$.
In that case, $c^{(1)}_k(\bm{z})$ must be updated as follows:
\begin{align}
    c^{(1)}_k(\bm{z}) &\to c^{(1)}_k(\bm{z}) + 2J_{s,s} \Delta z_s, \quad &&\text{for } k = s, \nonumber \\
    c^{(1)}_k(\bm{z}) &\to c^{(1)}_k(\bm{z}) + J_{s,k} \Delta z_s, \quad &&\text{for } k \neq s.
\end{align}

Next, we consider the case of HUIO [Eq. (\ref{def_HUIO})].
To calculate the energy difference, we extract from $E_p(\bm{z})$ the terms that involve the variable $z_k$:
\begin{align}
T_k(\bm{z}) = \sum_{m\in \Omega_k} a^{(m)}_{k} z_k^{m},
\end{align}
with
\begin{align}
    a^{(m)}_{k}(\bm{z}) &= J_{\underbrace{k, \dots, k}_{m\text{ times}}}
    +\sum_{\substack{i_1 \\ i_1 \neq k}} J_{i_1,\underbrace{k, \dots, k}_{m\text{ times}}} z_{i_1}\nonumber \\ 
    &+\sum_{\substack{i_1, i_2 \\ i_1 \neq k, i_2 \neq k}} J_{i_1, i_2,\underbrace{k, \dots, k}_{m\text{ times}}} z_{i_1} z_{i_2}
    +\cdots.
\end{align}
Here, $a^{(m)}_{k}(\bm{z})$ is the coefficient of the monomial $z_k^{m}$, and it depends on all variables other than $z_k$.
The set $\Omega_k$ denotes the exponents of $z_k$ that appear in the objective function $E(\bm{z})$.
The contribution to the energy difference arises only from $T_k(\bm{z})$, so that
\begin{align}
\Delta E_{p,k}(\bm{z})
    &= T_k(\bm{z} + \Delta \bm{z}_k) - T_k(\bm{z}) \nonumber\\
    &= \sum_{m\in \Omega_k} a^{(m)}_{k}(\bm{z}) \left\{ \left(z_k + \Delta z_k\right)^m - z_k^m \right\}.
    \label{energy_diff_poly}
\end{align}
An important point here as well is that the coefficients $a^{(m)}_{k}(\bm{z})$ are independent of the change $\Delta z_k$.
Therefore, once these coefficients are precomputed for all variables, the energy difference can be obtained in essentially $O(|\Omega_k|)$ time using Eq. (\ref{energy_diff_poly}), assuming that the $m$-th power can be evaluated in $O(1)$.
As in the QUIO case, one could expand Eq. (\ref{energy_diff_poly}) in terms of $\Delta z_k$ and store all resulting coefficients.
However, storing these coefficients leads to a considerably more complicated implementation for HUIO, so it is more practical to retain the coefficients $a^{(m)}_{k}$ directly.
In addition, in many practical instances $|\Omega_k|$ is relatively small, and evaluating the energy difference using Eq. (\ref{energy_diff_poly}) is sufficiently fast.

These coefficients must, of course, be updated when a variable is updated.
Suppose that $z_s$ is updated to $z_s + \Delta z_s$.
Since $a^{(m)}_{k}(\bm{z})$ does not depend on the variable $z_k$, only the coefficients $a^{(m)}_{k}(\bm{z})$ with $k \neq s$ need to be updated.
To evaluate this update, as before, we extract from $a^{(m)}_{k}(\bm{z})$ only the terms that contain $z_s$, which can be written as follows.  
\begin{align}
    U^{(m)}_{k,s}(\bm{z})=\sum_{n\in \Omega_{k,s}} b^{(m,n)}_{k,s}(\bm{z}) z^{n}_s,
\end{align}
with
\begin{align}
    b^{(m,n)}_{k,s}(\bm{z})
    =&J_{\underbrace{s, \dots, s}_{n\text{ times}},\underbrace{k, \dots, k}_{m\text{ times}}} \nonumber\\
    &+\sum_{\substack{i_1 \\ i_1 \neq k\neq s}} J_{i_1,\underbrace{s, \dots, s}_{n\text{ times}},\underbrace{k, \dots, k}_{m\text{ times}}} z_{i_1} \nonumber\\
    &+\sum_{\substack{i_1,i_2 \\ i_1 \neq k\neq s,i_2 \neq k\neq s}} J_{i_1,i_2,\underbrace{s, \dots, s}_{n\text{ times}},\underbrace{k, \dots, k}_{m\text{ times}}} z_{i_1} z_{i_2} + \cdots.
\end{align}
In this expression, $b^{(m,n)}_{k,s}(\bm{z})$ represents the coefficient of the monomial $z_s^{n}$ in $a^{(m)}_{k}(\bm{z})$, and it depends on all variables other than $z_s$ and $z_k$.
The set $\Omega_{k,s}$ denotes the exponents of $z_s$ in those terms of the objective function $E_p(\bm{z})$ that contain both $z_k$ and $z_s$.  
By using $b^{(m,n)}_{k,s}(\bm{z})$, the coefficients $a^{(m)}_{k}(\bm{z})$ must be updated as follows.
\begin{align}
    a^{(m)}_{k}(\bm{z})\to a^{(m)}_{k}(\bm{z}) + \sum_{n\in \Omega_{k,s}} b^{(m,n)}_{k,s}(\bm{z}) \{(z_s + \Delta z_s)^n - z^{n}_s\}
    \label{energy_diff_upd}
\end{align}

\subsection{Transition probability}\label{subsect-transi-prob}
In this subsection, we describe the transition probability $P(\Delta E_{p, k}, T)$ employed in this study.  
We first present two widely used transition probabilities, the Metropolis method and the heat bath method.  
We then introduce the optimal-transition Metropolis method.
\subsubsection{Metropolis method}\label{subsect-transi-prob-metro}
Let us first explain the Metropolis method, which is the simplest and most widely used method.
Assuming that the variable $z_k$ is chosen to be updated, its possible variable differences are
\begin{align}
    \Delta z^{(i)}_k = l_k - z_k + i,\quad i=0,1,\ldots,u_k-l_k.
\end{align}
Here, $\Delta z^{(i)}_k$ is the $i$-th possible change of $z_k$.
In the Metropolis method, a random nonzero change $\Delta z^{(i)}_k$ is first chosen for the variable $z_k$.
The transition probability from the current state to the proposed state
$z_k \to z_k + \Delta z^{(i)}_k$ is then given by
\begin{align}
    P(\Delta E^{(i)}_k, T)
      = \frac{1}{u_k - l_k}\min\left[1, \exp\left(-\frac{\Delta E^{(i)}_k}{T}\right)\right].
\end{align}
Here $\Delta E^{(i)}_k$ is the energy difference of the change $z_k \to z_k + \Delta z^{(i)}_k$ and $T$ is the current temperature.
The prefactor $1/(u_k - l_k)$ accounts for the uniform random choice among the possible nonzero changes.

The advantage of this update method is that the determination of the variable change does not depend on the number of possible changes of the variable, $u_k - l_k + 1$.
In QUIO and HUIO problems arising in mathematical optimization, variables may have very wide range; nevertheless, the transition probability can still be evaluated in $O(1)$ time.
On the other hand, since the variable difference is chosen at random,
the probability of selecting the optimal update for a single variable $z_k$
is $O(1/(u_k - l_k))$, and becomes very small when the variable ranges are wide.

\subsubsection{Heat bath method}
Next, we describe the heat bath method, which is another widely used method.
In this method, the probability of taking the update with energy difference $\Delta E^{(i)}_k$ for $z_k \to z_k + \Delta z^{(i)}_k$ is given by
\begin{align}
    P(\Delta E^{(i)}_k, T)=\frac{\displaystyle\exp\left(-\frac{\Delta E^{(i)}_k}{T}\right)}{\displaystyle\sum^{u_k - l_k}_{j=0}\exp\left(-\frac{\Delta E^{(j)}_k}{T}\right)}.
    \label{prob_heatbath}
\end{align}
Unlike the Metropolis method, the variable difference is not chosen at random. 
Instead, the probability is determined directly by the corresponding energy differences $\Delta E^{(i)}_k$.
For this reason, the heat bath method is often considered more efficient than the Metropolis method. 
However, it is important to note that in the heat bath method one must compute the transition probability for all possible changes $\Delta z^{(i)}_k$ in general.
As a result, the computational cost scales as $O(u_k - l_k)$, which can become prohibitive when the variable ranges are very wide, significantly increasing the overall runtime of SA.

In the special case that the HUIO contains only multilinear terms, the variable update can be determined in $O(1)$ time under the heat bath method.
Consider the following multilinear objective function:
\begin{align}
    E_p(\bm{z})
    =\sum_{i_1}J_{i_1} z_{i_1}+
    \cdots+\sum_{\substack{i_1,i_2,\ldots,i_p \\ i_1 \neq i_2\neq \cdots\neq i_p}}J_{i_1,i_2,\ldots,i_p} z_{i_1} z_{i_2}\cdots z_{i_p}.
\end{align}
In this case, the energy difference $\Delta E^{(i)}_k$ is linear in the variable change $\Delta z^{(i)}_k$:
\begin{align}
    \Delta E^{(i)}_k=a_k(\bm{z})\Delta z^{(i)}_k=a_k(\bm{z})(l_k - z_k + i),
\end{align}
with
\begin{align}
    a_k(\bm{z})=J_k + \sum_{\substack{i_1 \\ i_1 \neq k}}J_{i_1,k}z_{i_1} + \cdots + \sum_{\substack{i_1,\ldots,i_{p-1} \\ i_1 \neq \cdots \neq i_{p-1} \neq k}}J_{i_1,\ldots,i_{p-1},k}z_{i_1}\cdots z_{i_{p-1}}.
\end{align}
Accordingly, the cumulative transition probability can be written as
\begin{align}
    P_{\text{sum}}(\Delta E^{(i)}_k, T) &= \sum^{i}_{j=0}P(\Delta E^{(j)}_k, T)\nonumber\\
    &=\frac{1 - \displaystyle\exp\left(-\frac{a_k(\bm{z})(\Delta z^{(i)}_k + z_k - l_k + 1)}{T}\right)}{1 - \displaystyle\exp\left(-\frac{a_k(\bm{z})(u_k - l_k + 1)}{T}\right)}.
\end{align}
By using the inverse transform sampling\cite{Devroye1986}, we can obtain $\Delta z_k$ according to the probability in Eq. (\ref{prob_heatbath}):
\begin{align}
    \Delta z_k = \lceil P^{-1}_{\text{sum}}(u, T) \rceil,
\end{align}
with
\begin{align}
    &P^{-1}_{\text{sum}}(u, T) \nonumber \\
    &=- \frac{T}{a_k(\bm{z})} \ln\left[1 - u \left\{1 - \exp\left(-\frac{a_k(\bm{z})(u_k - l_k + 1)}{T}\right)\right\}\right] \nonumber \\
     &\quad\;- (z_k - l_k + 1).
    \label{inverse_prob}
\end{align}
Here, $u$ is a random real number sampled uniformly from the interval $[0, 1]$.
Importantly, Eq. (\ref{inverse_prob}) can be evaluated in $O(1)$ time, so that the variable change under the heat bath method is also determined in $O(1)$ time in this case.

\subsubsection{Optimal-transition Metropolis method}
As summarized above, the Metropolis method has low computational cost but suffers from a low probability of moving in a favorable direction due to its random proposal mechanism.  
In contrast, the heat bath method provides more informed updates but requires higher computational cost, except in special cases.  
To overcome these difficulties, we introduce another method, which we call the optimal-transition Metropolis method in this paper.
The idea of our method is to modify the proposal step of the Metropolis method by introducing the optimal transition with a certain probability $0 \leq r \leq 1$.
Here, the optimal transition refers to the variable change $\Delta z_k$ that yields the minimum energy difference in Eq. (\ref{energy_diff_poly}):
\begin{align}
    \Delta z^{\min}_k = \arg\min_{\Delta z_k}\,\Delta E_{p,k}.
\end{align}
Note that, if this transition were always chosen, the system would be trapped in a local minimum.  
To avoid this, we introduce a parameter $r = s/(N_\text{sweep}-1)$ with $s=0,1,\ldots,N_{\text{sweep}}-1$, defined by the progress of the annealing schedule, which controls how often the optimal transition is attempted.
Namely, the probability that a candidate change $\Delta z_k^{(i)}$ is accepted is given by
\begin{align}
    &P(\Delta E^{(i)}_k) = \frac{1-r}{u_k - l_k}\min\left[1, \exp\left(-\frac{\Delta E^{(i)}_k}{T}\right)\right], \text{ for }\Delta E^{(i)}_k \neq \Delta E^{\min}_k,\\
    &P(\Delta E^{\min}_k) = r\min\left[1, \exp\left(-\frac{\Delta E^{\min}_k}{T}\right)\right].
    \label{eq_opt_trans}
\end{align}
Here, $\Delta E^{\min}_k=\min_{\Delta z_k}\,\Delta E_{p,k}$ is the minimum energy difference among all possible changes of $z_k$.
We also note that similar attempts to improve proposal mechanisms have been made in SA for continuous variables.\cite{Dekkers1991, Ali1997, ALI200287, Locatelli2002}
In general, obtaining $\Delta z^{\min}_k$ requires minimizing a polynomial expression of the energy difference with respect to $\Delta z_k$, which can be computationally expensive.  
However, when the polynomial degree is small, the minimum can be found efficiently.  
In this study, we implemented the optimal-transition Metropolis method in cases where the energy difference is at most a fourth-order polynomial in $\Delta z_k$, which corresponds to HUIO problems in which each variable appears with degree at most four within each monomial.

It should be noted that, unlike the Metropolis and heat bath methods, the optimal-transition Metropolis method does not satisfy the balance condition for the Boltzmann distribution even when the temperature $T$ is kept fixed.
Therefore, it cannot be used for equilibrium sampling at a fixed temperature and can be employed only as a heuristic for searching low-energy states within SA.

\subsection{Initial and final temperature}\label{subsect-init-temp}
In this subsection, we explain how the initial and final temperatures in SA are determined.\cite{10.1063/1.34823, Aarts1985StatisticalC}
The initial temperature must be sufficiently high for the system.
If the initial temperature is set too low, the system easily becomes trapped in local minima.
If it is set too high, the high-temperature part of the search becomes essentially wasteful, and when the total annealing time is fixed, the low-temperature phase is effectively shortened, which again degrades the solution quality.
Thus, the initial temperature should be chosen to be sufficiently high but not excessive.

The final temperature must likewise be sufficiently low for the system.
If it remains too high, the energy cannot decrease enough, and the quality of the final solution is also degraded.
If it is too low, on the other hand, state updates rarely occur and computation time is wasted without further improvement, although setting it too low is still preferable to setting it too high.

An important point is that the appropriate initial and final temperatures cannot be determined solely from the properties of the model.
They also depend on other factors such as the total annealing time, the update scheme of the states, and the specific temperature schedule.
However, it is quite difficult to determine these temperatures in a way that also accounts for these factors.
In this study, therefore, we determine both temperatures solely based on the model information.
To ensure robustness against variations in the annealing time or cooling schedule, the initial temperature is slightly overestimated on the high-temperature side, and the final temperature is slightly underestimated on the low-temperature side.

Let us now discuss how the initial temperature is determined in practice.
The key point is the maximum energy difference that can occur when a variable is updated.
In general, at high temperatures, it is necessary to allow transitions that increase the energy so that the system can escape from local minima.
Therefore, it is desirable that even the largest possible energy increase, denoted by $\Delta E_{\max}$, can be accepted with a sufficiently large probability.
Considering the Metropolis method, in which the transition probability takes a simple exponential form, the temperature at which a transition with an energy increase of $\Delta E_{\max}$ occurs with probability $a$ is given by
\begin{align}
T_{\text{init}} = -\frac{\Delta E_\text{max}}{\ln a}.
\label{eq_temp_init}
\end{align}
In this study, we set $a = 1/2$.
Since it is generally difficult to evaluate $\Delta E_{\max}$ exactly, we approximate it using the following expression derived from Eq. (\ref{energy_diff_poly}):
\begin{align}
\Delta E_{\max} \simeq \max_{k} \sum_{m\in\Omega_k} \tilde{a}^{(m)}_k(\bm{z})(u_k - l_k)^m,
\end{align}
with
\begin{align}
\tilde{a}^{(m)}_k(\bm{z})&= \|J_{\underbrace{k, \dots, k}_{m\text{ times}}}\|
    +\sum_{\substack{i_1 \\ i_1 \neq k}} \|J_{i_1,\underbrace{k, \dots, k}_{m\text{ times}}}\| z^{\text{max}}_{i_1}\nonumber \\ 
    &+\sum_{\substack{i_1, i_2 \\ i_1 \neq k, i_2 \neq k}} \|J_{i_1, i_2,\underbrace{k, \dots, k}_{m\text{ times}}}\| z^{\text{max}}_{i_1} z^{\text{max}}_{i_2}
    +\cdots,
\end{align}
and
\begin{align}
    z^{\text{max}}_i = \max(\|l_i\|, \|u_i\|).
\end{align}

Finally, let us consider the determination of the final temperature.
Even at sufficiently low temperatures in SA, it is often beneficial to accept, with a small probability, transitions that slightly increase the energy, since this allows the system to escape from local minima.
In this case, the key quantity is the minimum possible energy increase, denoted by $\Delta E_{\min}$.
The temperature at which a transition with this energy difference is accepted with a very small probability $b$ is given by
\begin{align}
T_{\text{final}} = -\frac{\Delta E_{\min}}{\ln b}.
\end{align}
In this study, we set $b = 0.001$.
As it is also difficult to evaluate $\Delta E_{\min}$ exactly, we approximate it by using the smallest absolute value among all interaction coefficients, as follows:
\begin{align}
    \Delta E_{\min} \simeq \min\left(
  \min_{i_1} |J_{i_1}|,
  \min_{i_1,i_2} |J_{i_1,i_2}|,
  \ldots,
  \min_{i_1,\ldots,i_p} |J_{i_1,\ldots,i_p}|
\right).
\end{align}
These definitions of the initial and final temperatures provide a practical way to set appropriate temperature ranges for SA based only on the properties of the objective function, and we use them in the same manner for all transition probabilities considered in this study.

\section{Results}\label{sect-results}
In this section, we present numerical results for SA with the transition probabilities described in Sect.~\ref{subsect-transi-prob}.
Our proposed algorithms are implemented in the open-source library OpenJij\cite{openjij}, and are freely available for use.
All numerical experiments presented below were performed using OpenJij on a machine equipped with an Apple M1 Max processor and 64 GB of memory.
In addition, we used the libraries JijModeling\cite{cite_jijmodeling} and OMMX\cite{cite_ommx_1, cite_ommx_2} to implement the QUIO and HUIO problems.
Both libraries provide Python front ends with Rust back ends, enabling fast construction of problem instances.

We evaluate the algorithms on the following benchmark models:
\begin{align}
    E^{\text{FC}}_{p, u}(\bm{z}) &= -\frac{1}{u^2}\sum_{i=1}^{N}\sum_{j=i+1}^{N} z_i z_j -\frac{1}{u^p} \sum_{i=1}^{N} z_i^{p}, \quad -u \leq z_i \leq u,
    \label{model-FC}
\end{align}
\begin{align}
     E^{\text{ML}}_{p, u}(\bm{z}) &= -\frac{1}{u^p}\sum_{i=1}^{N} z_{i} z_{i + 1} \cdots z_{i + p - 1}, \quad -u \leq z_i \leq u,
    \label{model-ML}
\end{align}
\begin{align}
    E^{\text{RI}}_{p, u}(\bm{z}) &= \frac{1}{u^p}\sum_{i_1,\ldots,i_p}J_{i_1,\ldots,i_p} z_{i_1} \cdots z_{i_p}, \quad -u \leq z_i \leq u.
    \label{model-RI}
\end{align}
Here, $N$ is the number of variables and we set $N=100$ in the following experiments.
We refer to the models $E^{\text{FC}}_{p,u}(\bm{z})$, $E^{\text{ML}}_{p,u}(\bm{z})$, and $E^{\text{RI}}_{p,u}(\bm{z})$ as the fully connected model, the multilinear model, and the random interaction model, respectively.
{\color{red}
For the multilinear model, we impose periodic boundary conditions, $z_{i+N}=z_i$.
For the random interaction model, each interaction term is generated by selecting
$p$ indices $i_1,\ldots,i_p$ with repetition allowed and assigning
a coefficient $J_{i_1,\ldots,i_p}$ sampled uniformly from $[-1,1]$;
hence on-site terms such as $z_i^k$ can appear.
In the following experiments, we set the number of interaction terms to 1000 and use 100 independent random realizations.
}

{\color{red}
We benchmark these models from the following viewpoints.
The fully connected model provides a dense interaction structure with a large number of terms, leading to a high computational cost, while its ground-state energy is analytically known to be $-N(N+1)/2$.
The multilinear model also has an analytically known ground-state energy, $-N$; however, for large $p$, it becomes extremely difficult to reach the ground state by single-variable updates, since an interaction term vanishes if any one of its variables takes the value zero, resulting in a large number of configurations with nearly flat energy contributions.
Finally, the random interaction model is sparse and has a relatively small number of interaction terms, but its ground state is nontrivial.
}

The remainder of this section is organized as follows.
In Sect.~\ref{sect_time}, we analyze how {\color{red}the annealing time depends on the model parameters $u$ and $p$ for a fixed number of sweeps.
}
In Sect.~\ref{sect_temperature}, we verify that the initial and final temperatures used in SA, as described in Sect.~\ref{subsect-init-temp}, are appropriate for these models.
In Sect.~\ref{sect_sweeps}, we examine how the energy decreases as a function of the number of sweeps.
In Sect.~\ref{sect_energy}, we evaluate the performance of SA on these models in terms of both solution quality and annealing time.

\subsection{Annealing time under fixed sweeps}\label{sect_time}
In this subsection, {\color{red}we focus on the annealing time of SA for a fixed number of sweeps,
where the annealing time is the elapsed wall-clock time required to complete a single SA run.
By fixing the number of sweeps, we use the annealing time to estimate the computational cost of each algorithm and examine its dependence on the model parameters.}
In the following numerical experiments, SA was performed with 100 samples and 1000 sweeps, and the mean annealing time per single SA run was computed.

\begin{figure}[!t]
  \centering
  \includegraphics[width=\columnwidth]{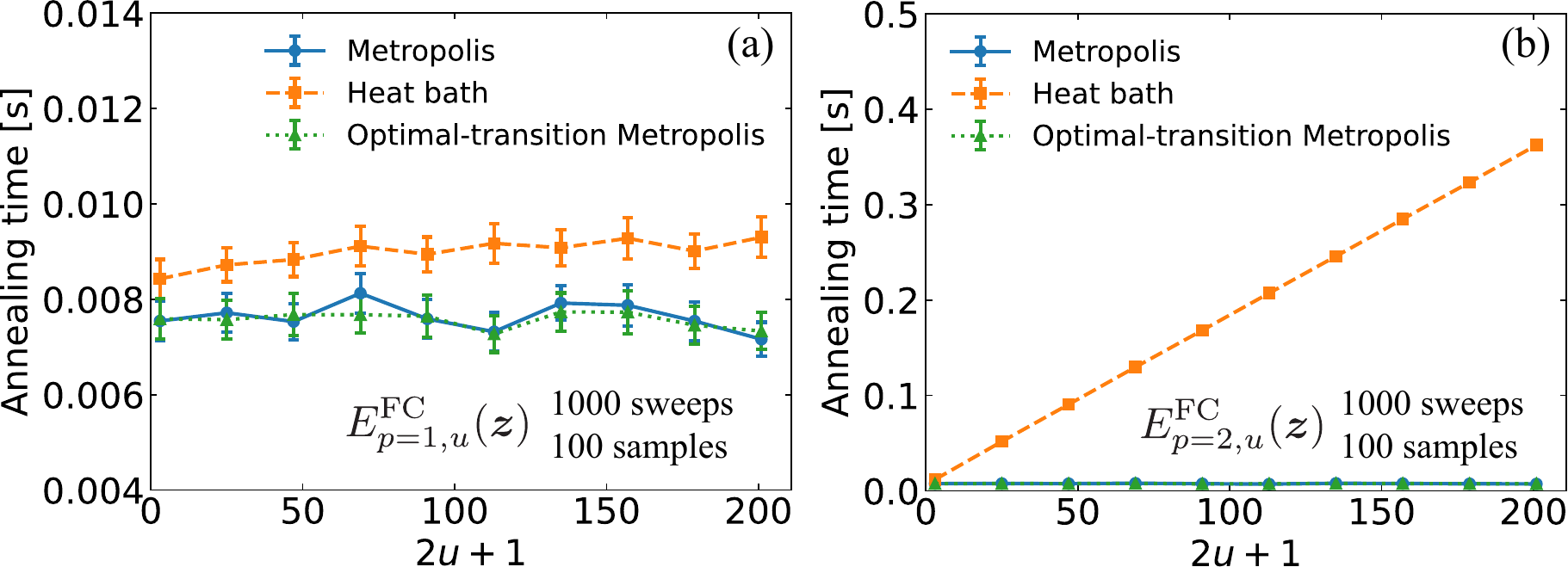}
  \caption{
    Annealing time as a function of variable range $2u + 1$ for
    (a) $E^{\text{FC}}_{p=1,u}(\bm{z})$ and
    (b) $E^{\text{FC}}_{p=2,u}(\bm{z})$.
    Each data point represents the mean annealing time over 100 samples, with 1000 sweeps and $N = 100$ variables.
    Error bars indicate the standard errors.
  }
  \label{time-bound-dep}
\end{figure}

Let us first investigate how the variable range, $2u + 1$, influences the annealing time,  
focusing on the fully connected model [Eq. (\ref{model-FC})].  
Figure \ref{time-bound-dep} shows the annealing time as a function of $2u + 1$, for $p = 1$ and $p = 2$.  
For $p = 1$, the annealing time does not increase with $2u + 1$ for any of the transition probabilities [Fig. \ref{time-bound-dep}(a)], because determining the update of a single variable does not depend on the variable range for all transition methods.
For $p = 2$, only the heat bath method requires $O(2u + 1)$ time to determine the update of a single variable,  
because all $2u + 1$ candidate values must be examined when performing the update.  
Consequently, as shown in Fig. \ref{time-bound-dep}(b), only the heat bath method exhibits annealing time that increases linearly with $2u + 1$.
\begin{figure}[!t]
  \centering
  \includegraphics[width=\columnwidth]{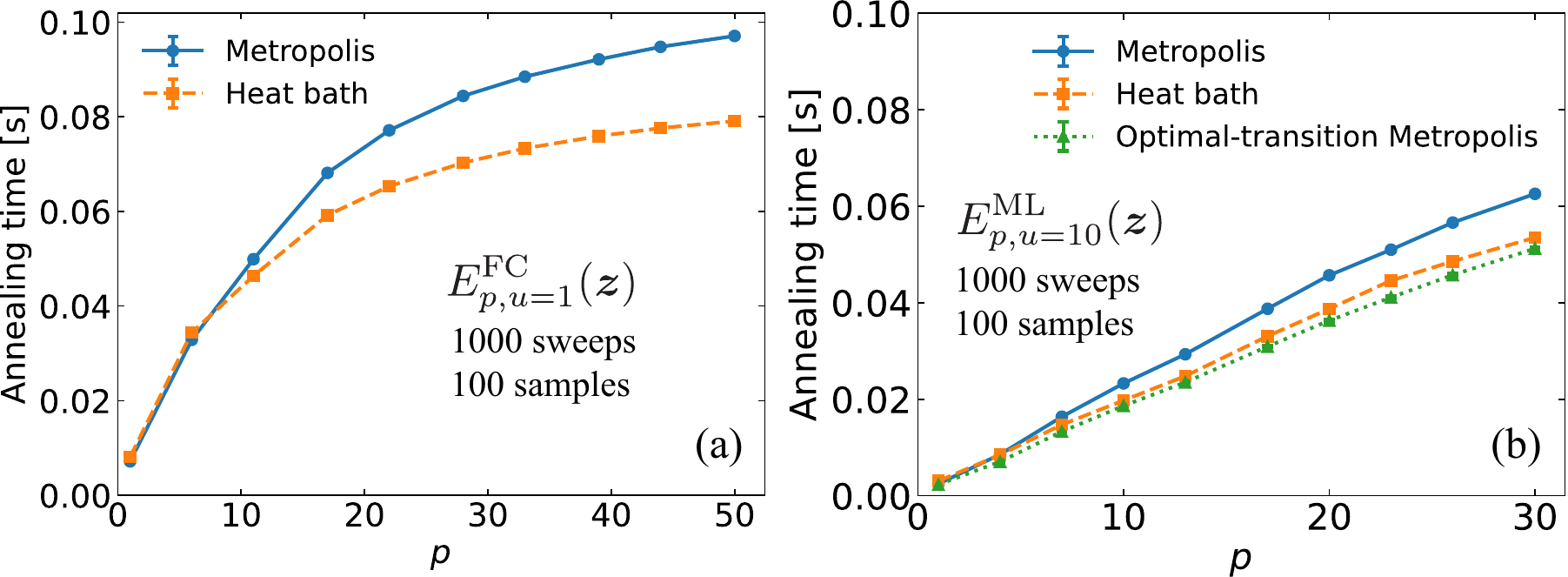}
  \caption{
  Annealing time as a function of $p$ for
    (a) $E^{\text{FC}}_{p,u=1}(\bm{z})$ and
    (b) $E^{\text{ML}}_{p,u=10}(\bm{z})$.
    Each data point represents the mean annealing time over 100 samples, with 1000 sweeps and $N = 100$ variables.
    Error bars indicate the standard errors but are too small to be visible.
  }
  \label{time-p-dep}
\end{figure}

Next, we investigate how the exponent $p$ in the monomial $z^p$ influences the annealing time.
Figure \ref{time-p-dep}(a) shows the annealing time of SA for the fully connected model $E^{\text{FC}}_{p,u=1}(\bm{z})$ as the degree $p$ is varied.
The optimal-transition Metropolis method is not included in this plot, as it is implemented only for cases where $p \leq 4$.
One can see that the annealing time increases slightly as $p$ becomes larger.  
In general, the dominant computational cost in SA arises from calculating the transition probabilities and updating the coefficients used for computing the energy difference [Eq. (\ref{energy_diff_upd})].
In the fully connected model $E^{\text{FC}}_{p,u=1}(\bm{z})$, both of these operations require evaluating polynomials whose degree is at most $p$, which naturally leads to a slight increase in annealing time as $p$ grows.

Finally, we investigate how the size $p$ of the multilinear terms influences the annealing time, focusing on the multilinear model [Eq. (\ref{model-ML})].
Since this model does not contain any terms involving powers of variables,  
each variable can be updated in $O(1)$ time, regardless of $p$.  
On the other hand, each variable interacts with $p - 1$ other variables.  
Therefore, when a variable is updated, the coefficients associated with those $p - 1$ variables in the energy difference must be updated [Eq. (\ref{energy_diff_upd})].
As a result, the annealing time of this model is expected to increase linearly with $p$.  
Indeed, Fig. \ref{time-p-dep}(b) shows the annealing time of $E^{\text{ML}}_{p, u=10}(\bm{z})$ as a function of $p$, and we can confirm that, for all transition probabilities, the annealing time increases linearly with $p$.

In summary, most of the annealing time in SA is spent on two parts:  
the determination of variable updates according to the transition probabilities,  
and the update of the coefficients [Eq. (\ref{energy_diff_upd})] when a variable is actually updated.  
The former can be performed in $O(1)$ time for both the Metropolis and optimal-transition Metropolis methods,  
whereas for the heat bath method, the computation requires $O(u_k - l_k)$ time if the interactions include powers of variables.  
On the other hand, the latter part increases linearly with the number of variables that interact with the updated variable,  
regardless of the choice of transition probability.

\subsection{Energy as a function of temperature}\label{sect_temperature}
In this subsection, we examine the temperature dependence of the energy during the execution of SA.
This analysis allows us to verify whether the initial and final temperatures used in SA are appropriate.
We also investigate how the rate of energy decrease differs depending on the choice of transition probability.

\begin{figure}[!t]
  \centering
  \includegraphics[width=\columnwidth]{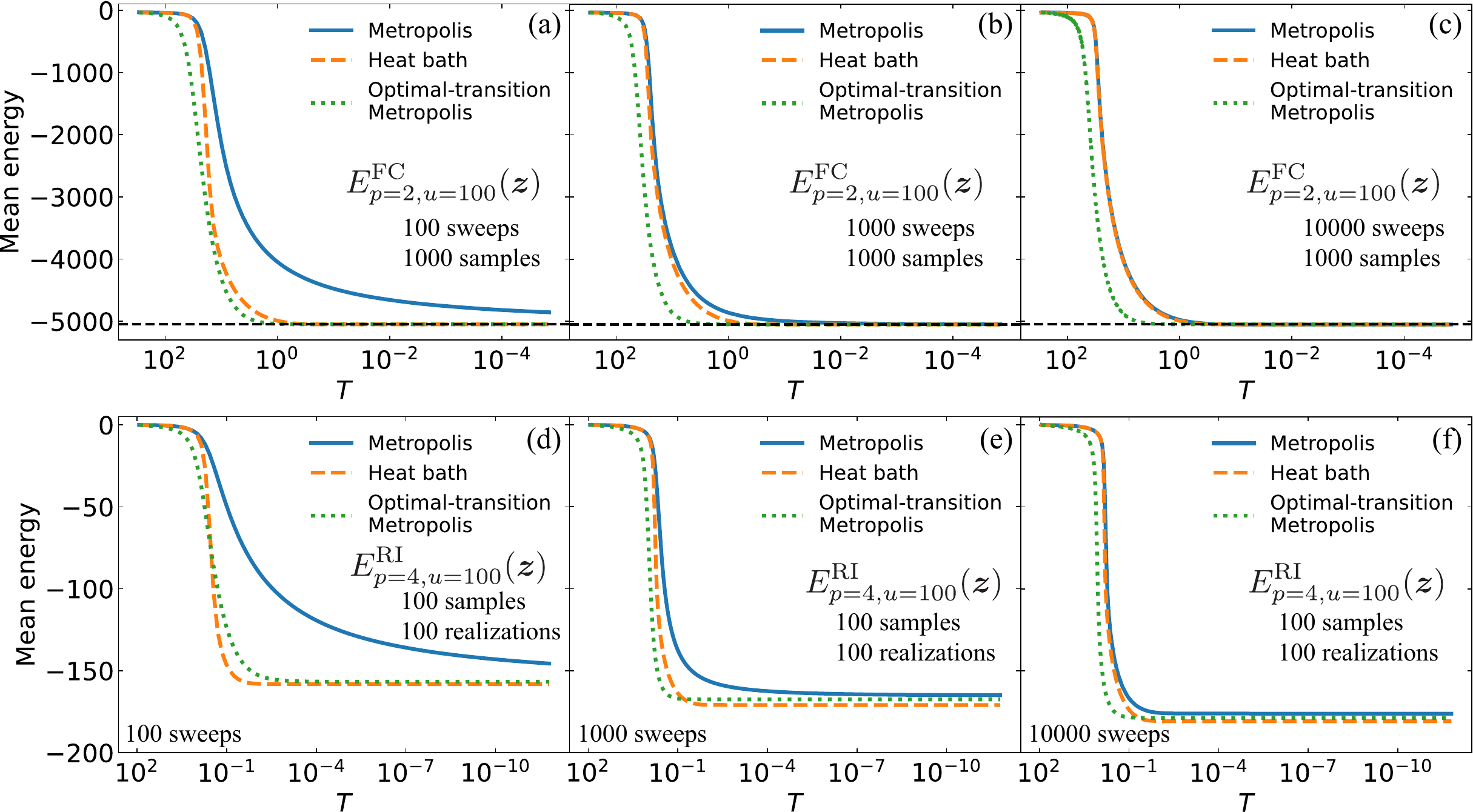}
  \caption{{\color{red}
    Temperature $T$ dependence of the mean energy during SA for 
    $E^{\text{FC}}_{p=2,u=100}(\bm{z})$ [(a)--(c)] and
    $E^{\text{RI}}_{p=4,u=100}(\bm{z})$ [(d)--(f)], with 100, 1000, and 10000 sweeps, respectively.
    Each data point is averaged over 1000 samples for $E^{\text{FC}}_{p,u}(\bm{z})$ and over 100 samples for each of 100 independent random realizations for $E^{\text{RI}}_{p,u}(\bm{z})$.
    The number of variables is $N = 100$ and the black dotted line in (a)--(c) marks the ground-state energy, $-5050$.
    }
  }
  \label{temperature-T-dep-general}
\end{figure}

{\color{red}
Figures~\ref{temperature-T-dep-general}(a)--(c) show the temperature dependence of the energy for
$E^{\text{FC}}_{p=2,u=100}(\bm{z})$ with 100, 1000, and 10000 sweeps, respectively.
Each data point is averaged over 1000 samples.
}
On the high-temperature side, there exists a narrow region in which the energy does not decrease and remains nearly constant.
Such a region indicates that the system is already at sufficiently high temperatures for these models, and it is desirable that the extent of this region be as small as possible.
Similarly, on the low-temperature side, there also exists a region where the energy ceases to decrease, which suggests that the system has reached a temperature sufficiently low for the model.
Overall, for the model $E^{\text{FC}}_{p=2,u=100}(\bm{z})$, the present initial temperature appears to be appropriate, whereas the final temperature is somewhat too low {\color{red}in most cases.
An exception is Metropolis with 100 sweeps [Fig.~\ref{temperature-T-dep-general}(a)], for which the final temperature appears to be somewhat high.
As mentioned in Sect.~\ref{subsect-init-temp}, the optimal initial and final temperatures depend on the number of sweeps.
}

{\color{red}
Similarly, Figs.~\ref{temperature-T-dep-general}(d)--(f) show the results for the random interaction model $E^{\text{RI}}_{p=4,u=100}(\bm{z})$ with 100, 1000, and 10000 sweeps, respectively.
Each data point is averaged over 100 samples for each of 100 independent random realizations.
}
This model is a HUIO problem with fourth-order interactions and random coupling coefficients, and its characteristics therefore differ from those of the fully connected model {\color{red}$E^{\text{FC}}_{p=2,u=100}(\bm{z})$}.
Nevertheless, the results exhibit similar tendencies to those of {\color{red}$E^{\text{FC}}_{p=2,u=100}(\bm{z})$}. 
The initial temperature appears appropriate, whereas the final temperature is also somewhat too low {\color{red}in most cases.
An exception is Metropolis with 100 sweeps [Fig.~\ref{temperature-T-dep-general}(d)], for which the final temperature appears to be somewhat high.
}


As for the differences among transition probabilities, the energy tends to decrease more readily when using the heat bath method than when using the Metropolis method, since it takes into account all possible variable updates instead of randomly selecting a single candidate.
For the optimal-transition Metropolis method, the energy occasionally drops sharply when an optimal transition actually occurs, and thus its energy is generally lower than that of {\color{red}the Metropolis method at the same temperature.
For the random interaction model, however, this method appears to become trapped in local minima at low temperatures, and its final energy ends up slightly higher than that of the heat bath method [Figs.~\ref{temperature-T-dep-general}(d)--(f)].
These differences become smaller as the number of sweeps increases.
}
\begin{figure}[!t]
  \centering
  \includegraphics[width=\columnwidth]{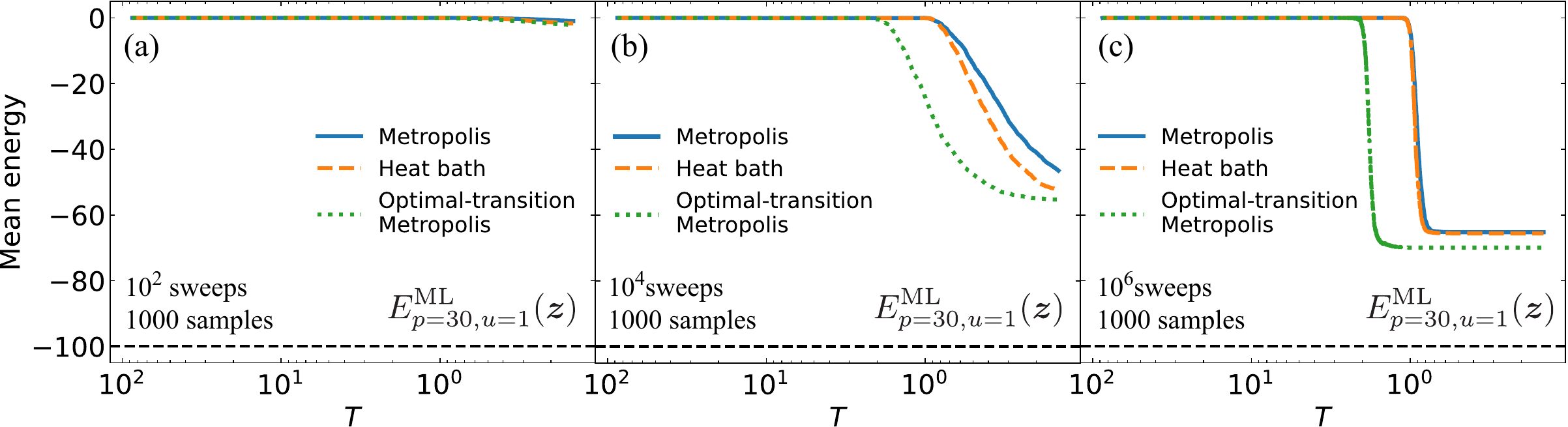}
  \caption{{\color{red}
    Temperature $T$ dependence of the mean energy during SA for 
    $E^{\text{ML}}_{p=30,u=1}(\bm{z})$, with (a) $10^2$, (b) $10^4$, and (c) $10^6$ sweeps.
    Each data point is averaged over 1000 samples.
    The number of variables is $N = 100$,
    and the black dotted line indicates the ground-state energy, $-100$.
    }
  }
  \label{temperature-T-dep-special}
\end{figure}

Next, we examine the results for the multilinear model $E^{\text{ML}}_{p=30,u=1}(\bm{z})$.
{\color{red}For $10^2$ and $10^4$ sweeps [Figs.~\ref{temperature-T-dep-special}(a) and \ref{temperature-T-dep-special}(b)]}, the energy is still decreasing even in the low-temperature region, suggesting that the final temperature is too high.
In contrast, for $10^6$ sweeps [Fig. \ref{temperature-T-dep-special}(c)], a region appears at low temperatures where the energy no longer decreases, indicating that the final temperature is sufficiently low.
These results show that while the initial temperature is sufficiently high, the final temperature is not always appropriate and depends on the number of sweeps.
In general, when the number of sweeps is small, the final temperature needs to be set lower.
With fewer sweeps, variable updates are limited, so allocating more sweeps to the low-temperature region becomes important.

{\color{red}
As discussed in the final part of Sect.\ref{sect-results}, for large $p$ the multilinear model exhibits a highly flat energy landscape, making it difficult to escape from such flat regions.
Indeed, even with $10^6$ sweeps, the energy does not reach the ground-state energy $-100$ [Fig.\ref{temperature-T-dep-special}(c)], indicating that many more sweeps are required.
}

In summary, the temperature-setting procedure in Sect. \ref{subsect-init-temp} is reasonable for the models studied here, 
although both the initial and final temperatures tend to be slightly conservative and are not necessarily optimal.
We note also that the final temperature can be too high when the number of sweeps is small.

\subsection{Energy as a function of sweeps}\label{sect_sweeps}
In this subsection, we examine how the energy decreases as a function of the number of sweeps.
Since the computational cost per sweep depends on the choice of transition probability, the energy as a function of sweeps does not by itself quantify performance, but it helps reveal the characteristic behavior of each transition probability.

\begin{figure}[!t]
  \centering
  \includegraphics[width=\columnwidth]{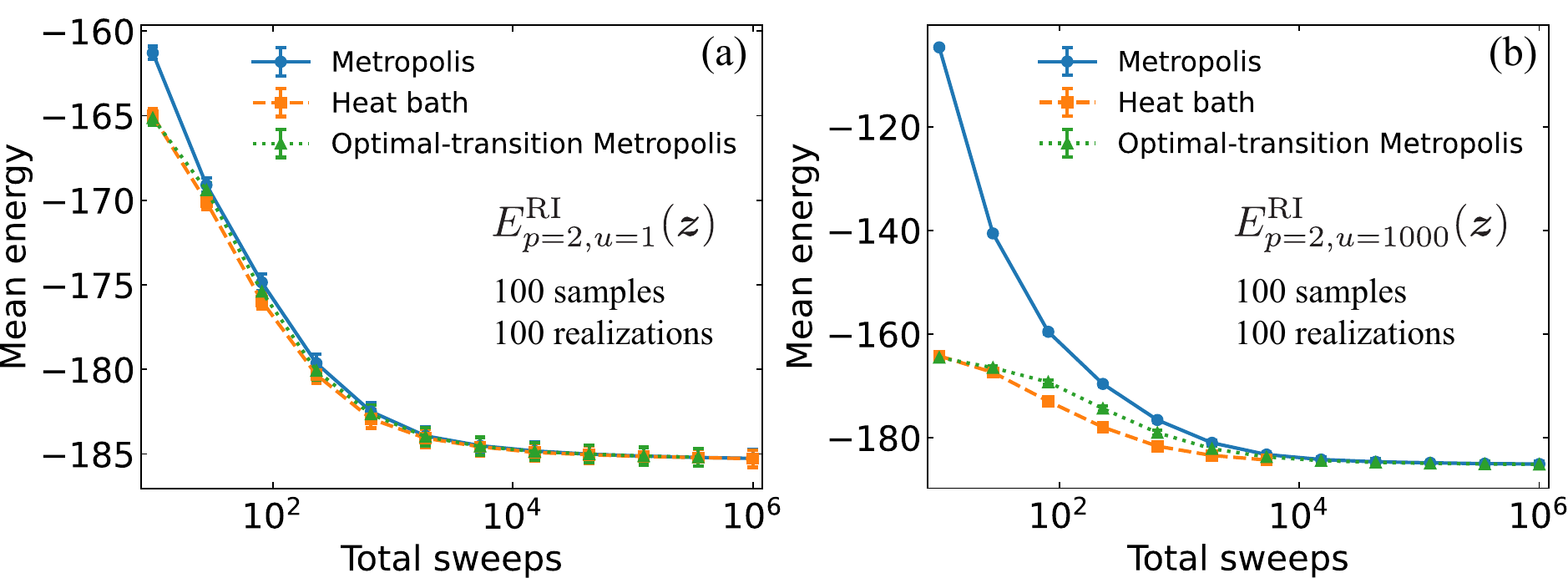}
  \caption{
  {\color{red}
  Sweep dependence of the mean energy for 
  (a) $E^{\text{RI}}_{p=2,u=1}(\bm{z})$ and
  (b) $E^{\text{RI}}_{p=2,u=1000}(\bm{z})$.
  Each data point represents the mean energy averaged over 100 samples for each of 100 independent random realizations.
  }
  Error bars indicate the standard errors but are too small to be visible.
  The number of variables is $N = 100$ and the number of interactions is 1000.
  }
  \label{sweep-dep-general}
\end{figure}

We first consider the random interaction model [Eq. (\ref{model-RI})] for $p = 2$.
{\color{red}The results are averaged over 100 samples for each of 100 independent random realizations.
}
When the variable range is small ($u = 1$), the three transition probability methods exhibit almost no difference in performance [Fig. \ref{sweep-dep-general}(a)]. 
In contrast, when the range is increased to {\color{red}$u = 1000$}, the Metropolis method yields noticeably higher energies, particularly at small sweep regimes [Fig. \ref{sweep-dep-general}(b)]. 
This behavior is consistent with the discussion in Sect. \ref{subsect-transi-prob-metro}, which states that the Metropolis method performs poorly when the variable range is wide.
It is also observed that the heat bath method achieves slightly lower energies than the optimal-transition Metropolis method in this regime [Fig. \ref{sweep-dep-general}(b)].

\begin{figure}[!t]
  \centering
  \includegraphics[width=\columnwidth]{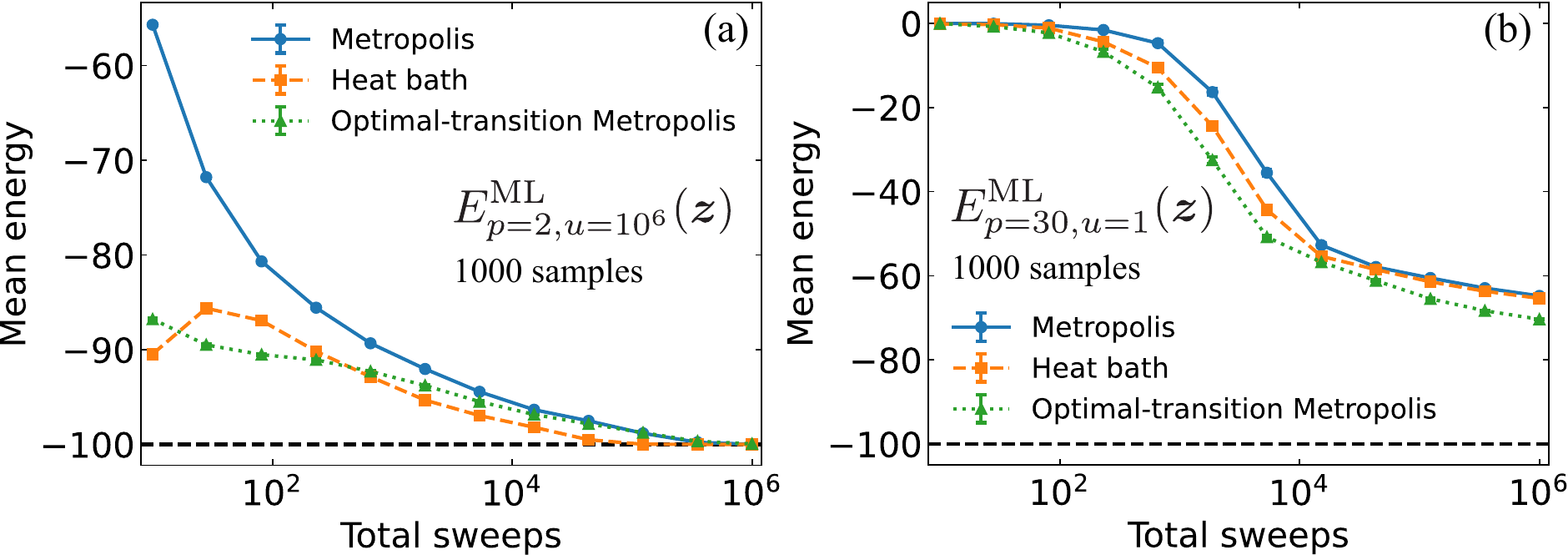}
  \caption{
  Sweep dependence of the mean energy for 
  (a) $E^{\text{ML}}_{p=2,u=10^6}(\bm{z})$ and 
  (b) $E^{\text{ML}}_{p=30,u=1}(\bm{z})$.
  Each data point represents the mean energy averaged over 1000 samples.
  Error bars indicate the standard errors but are too small to be visible.
  The number of variables is $N = 100$ and the black dotted line indicates the ground-state energy, $-100$.
  }
  \label{sweep-dep-special}
\end{figure}

The results for the multilinear model [Eq. (\ref{model-ML})]{\color{red}, averaged over 1000 samples,} show a somewhat different trend.
For the extremely large variable range ($u = 10^{6}$), the Metropolis method again performs poorly, but the heat bath method exhibits a temporary degradation at small sweeps [Fig. \ref{sweep-dep-special}(a)].
For the case of large polynomial degree ($p = 30$), the system does not reach the ground state even after $10^{6}$ sweeps for any transition probability, and in this regime the optimal-transition Metropolis method yields the lowest energies [Fig. \ref{sweep-dep-special}(b)].

In summary, when compared at the same number of sweeps,
the Metropolis method shows poor performance when the variable range is wide. For the other two methods, depending
on the number of sweeps, either the heat bath method or the optimal-transition Metropolis method can perform slightly
better, but the difference between them remains small.

\subsection{Energy as a function of annealing time}\label{sect_energy}
In this subsection, we examine the dependence of the optimized energy on the annealing time for different transition probabilities.
This analysis aims to evaluate the performance of each transition probability, where better performance corresponds to achieving a lower energy within a shorter annealing time.
In addition, we compare these results with those obtained by applying SA with the Metropolis method to QUBO and HUBO formulations that are constructed by converting the integer variables into binary ones using log encoding:
\begin{align}
z_k \to (u_k - l_k - 2^{m_k-1} + 1)b_{k,m_k-1} + l_k + \sum^{m_k-2}_{i=0}2^{i}b_{k,i}.
\end{align}
Here, $m_k=\lceil\log_2(u_k-l_k + 1)\rceil$ is the number of binary variables $b_{k,i}\in\{0, 1\}$ corresponding to an integer variable $z_k$.
Although the total computational cost should include the time required for binary encoding,
we exclude it here to focus on the dependence of the achieved energy on the annealing time.
It should be noted, however, that for models with higher-order interactions or large variable range,
such binary conversion can require a substantial amount of time.
{\color{red}In the following analysis, SA was performed for several fixed numbers of sweeps, using 1000 samples for the multilinear model and 100 samples for each of 100 independent random realizations for the random interaction model.
The mean energy obtained in each case was plotted against the mean annealing time.}

\begin{figure}[!t]
  \centering
  \includegraphics[width=\columnwidth]{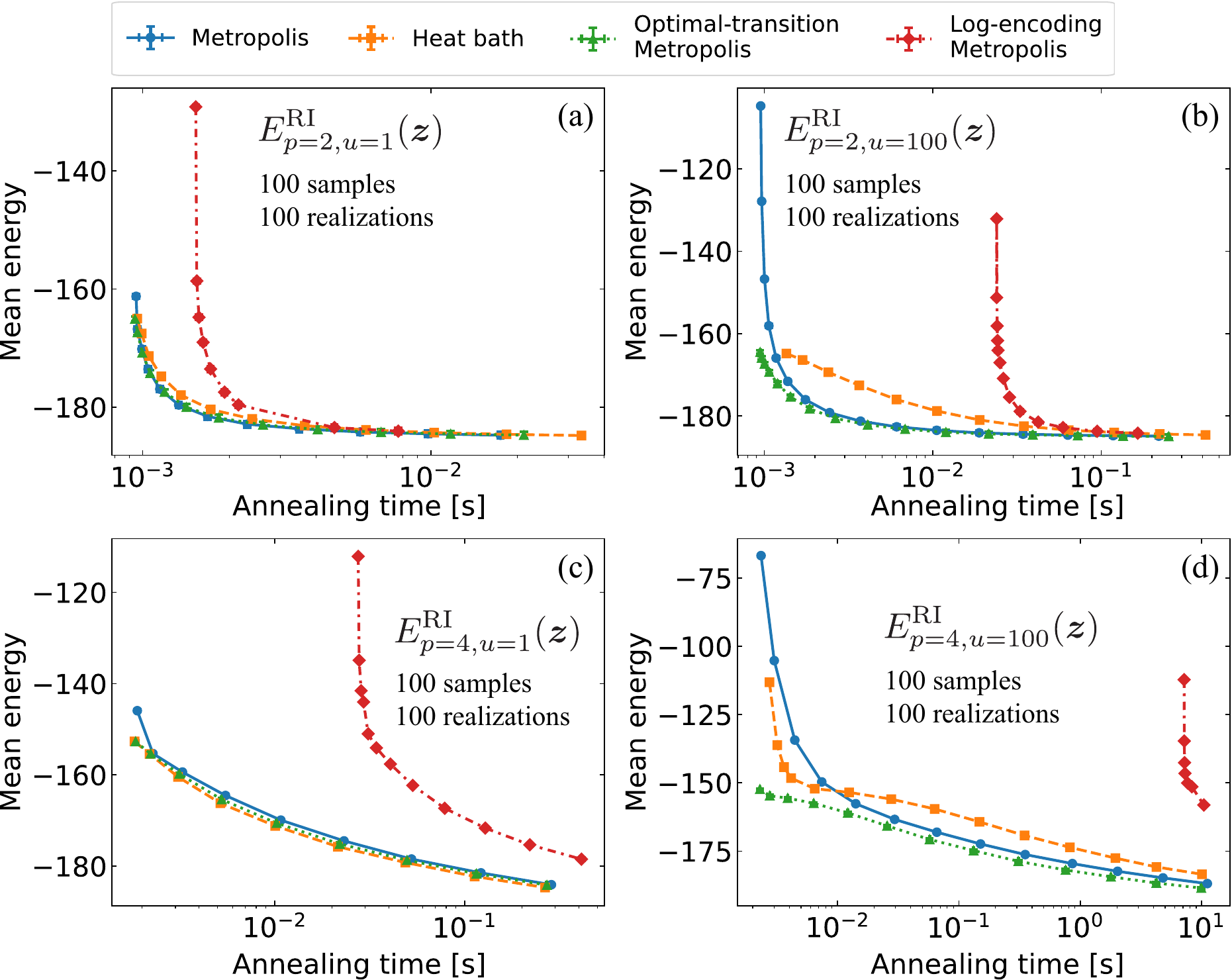}
  \caption{
  {\color{red}
  Annealing--time dependence of the mean energy for 
  (a) $E^{\text{RI}}_{p=2,u=1}(\bm{z})$,
  (b) $E^{\text{RI}}_{p=2,u=100}(\bm{z})$,
  (c) $E^{\text{RI}}_{p=4,u=1}(\bm{z})$, and
  (d) $E^{\text{RI}}_{p=4,u=100}(\bm{z})$.
  Each data point is averaged over 100 samples for each of 100 independent random realizations.}
  Error bars indicate the standard errors of both energy and annealing time but are too small to be visible.
  The number of variables is $N = 100$ and the number of interactions is 1000.
  }
  \label{time-dep-general}
\end{figure}

First, we examine the results for the random interaction model [Eq. (\ref{model-RI})].
Figure \ref{time-dep-general}(a) shows the results for $p=2$ and $u=1$.
When this model is converted into a QUBO formulation using the log encoding,  
the number of variables becomes twice as large and the number of interactions becomes four times larger,  
which increases the computational cost per sweep.
As a result, SA applied to the log-encoded QUBO formulation yields higher energies at short annealing times, indicating poorer solution quality.
In contrast, little difference is observed among the other transition probabilities, because when $u = 1$, each variable can take only three states ($-1, 0$, and $1$), so the influence of the choice of transition probability becomes less significant.
As the annealing time increases, the optimized energies obtained by all methods converge to similar levels, suggesting that all methods tend to reach the global optimum when sufficient annealing time is allowed.

Next, we examine the results for larger $u=100$ [Fig. \ref{time-dep-general}(b)].
In this case, the difference between the log-encoded QUBO formulation and the directly solved QUIO model becomes larger than in the case of $u = 1$.
This is simply because both the number of variables and the number of interaction terms become much larger than in the case of $u = 1$.
The heat bath method also exhibits slower energy convergence with respect to annealing time,
because approximately 1\% of the interaction terms are squared terms such as $z^2$,
and determining the updates for these variables requires $O(2u + 1)$ time in the heat bath method.
The optimal-transition Metropolis method shows lower energies in the short-time region 
and reaches the same final energies as the other methods in the long-time region, 
indicating that it provides the best overall performance.
This is due to the effect of the optimal transitions defined in Eq. (\ref{eq_opt_trans}).

Similar tendencies are also observed for the HUIO case.
As in the $p = 2$ case, a clear difference is observed between the log-encoded HUBO formulation and the directly solved HUIO model [Figs. \ref{time-dep-general}(c) and \ref{time-dep-general}(d)].
In general, when the model is log-encoded, the number of variables increases by a factor of $O(\log_2(2u+1))$, and each original $p$-body interaction term expands into $O((\log_2(2u+1))^p)$ binary terms. 
As a result, the computational cost per sweep increases substantially.
In particular, for $u = 100$, as shown in Fig.~\ref{time-dep-general}(d), the heat bath method again shows slower energy convergence, 
and the optimal-transition Metropolis method shows an even clearer advantage over the other methods.

\begin{figure}[!t]
  \centering
  \includegraphics[width=\columnwidth]{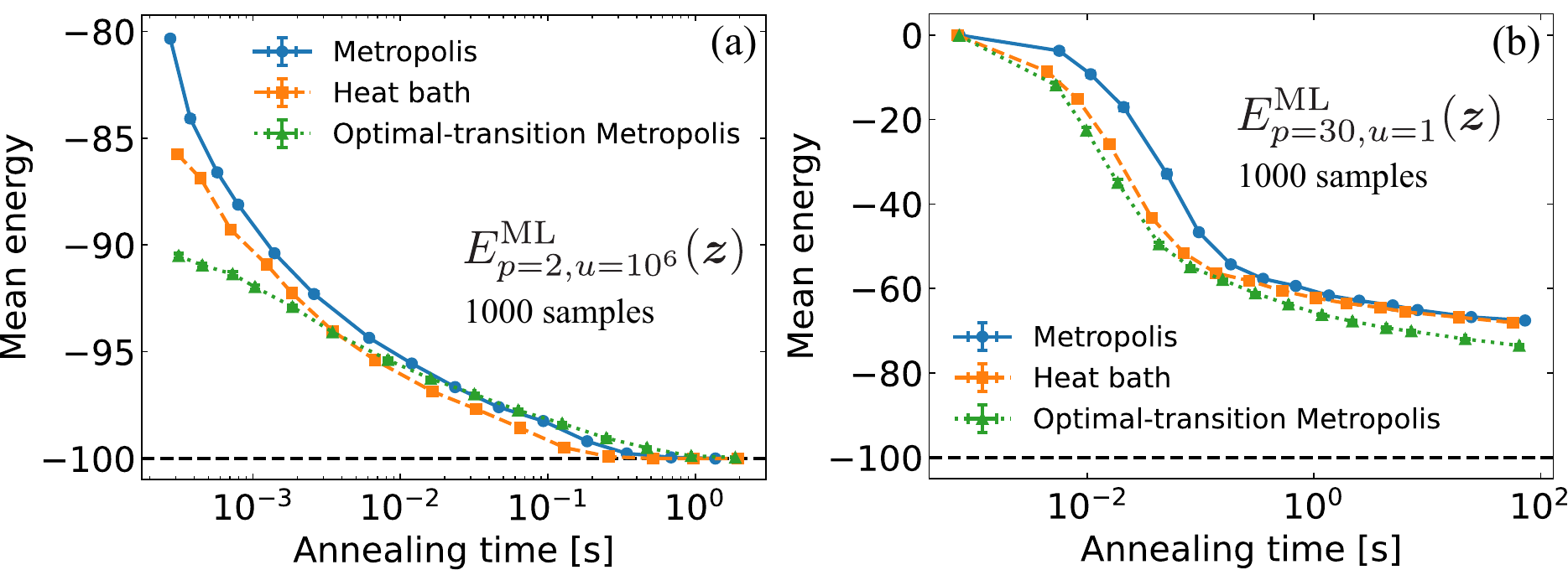}
  \caption{
  Annealing--time dependence of the mean energy obtained by SA with different transition probabilities for 
  (a) $E^{\text{ML}}_{p=2,u=10^6}(\bm{z})$ and 
  (b) $E^{\text{ML}}_{p=30,u=1}(\bm{z})$,
  Each data point represents the mean energy averaged over 1000 samples.
  Error bars indicate the standard errors of both energy and annealing time but are too small to be visible.
  The number of variables is $N = 100$ and the black dotted line indicates the ground-state energy, $-100$.
  }
  \label{time-dep-special}
\end{figure}

Finally, we examine the results for the multilinear model.
Note that, in this model, the heat bath method can determine the update of a variable in $O(1)$ time,
and thus the difference in computational time among the transition probabilities is small.
The log-encoded QUBO formulation is not plotted here, because the conversion process itself requires prohibitively long computation time.
Figure \ref{time-dep-special}(a) shows the results for $E^{\text{ML}}_{p=2,u=10^6}(\bm{z})$, where $u$ is extremely large.
For short annealing times, the optimal-transition Metropolis method achieves the lowest energies,
but it is overtaken by the heat bath method as the annealing time increases,
and eventually, all methods converge to the same energy for sufficiently long annealing times.
Figure \ref{time-dep-special}(b) shows the results for a larger polynomial degree, $p = 30, u=1$.
As discussed at the end of Sect.~\ref{sect_temperature}, this model contains many configurations for which multilinear terms take zero, so convergence to the optimal solution requires a very long annealing time.
Indeed, in Fig.~\ref{time-dep-special}(b), the energy does not reach the optimal value of $-100$, indicating that the annealing time is insufficient.
In this short-time regime, the optimal-transition Metropolis method achieves the lowest energies among all transition probabilities.
This tendency is also seen in {\color{red}Fig. \ref{temperature-T-dep-special}(c).}

In summary, directly solving the QUIO and HUIO is more effective than converting them into QUBO and HUBO by log encoding.  
The heat bath method becomes slow when the variable range is wide and high-order terms such as $z^p$ are present.  
Under these conditions, the optimal-transition Metropolis method shows better performance than the other transition probabilities, particularly for short annealing times.

\section{Summary}\label{sect-summary}
We proposed an SA method for optimizing polynomial functions composed of integer variables, namely the QUIO and HUIO problems.
By storing and updating the coefficients required for energy-difference computations, we can evaluate the transition probabilities efficiently.
We also proposed a method for estimating the initial and final temperatures from the characteristics of the model.
By comparing this direct application of SA with the traditional approach of converting the problems into QUBO or HUBO formulations, 
we showed that the direct approach achieves better performance in practice.

When the range of the integer variables are wide, the Metropolis method, which selects the next state randomly, has difficulty finding the optimal solution, 
while the heat bath method becomes time-consuming because it must evaluate all possible states of each variable.
To address these issues, we introduced the optimal-transition Metropolis method.
Comparing these transition probabilities, we found that all methods tend to reach the optimal solution when the annealing time is sufficiently long, 
but the optimal-transition Metropolis method shows a clear advantage over the others in the short-annealing-time regime.
This is particularly important in real-world applications, where the annealing time is often limited and the ability to obtain reasonably good solutions quickly is essential.
At the same time, this method does not lose effectiveness in the long-time regime, converging to the same solutions as the other transition probabilities.

In this study, the optimal-transition Metropolis method was implemented only in cases where each variable appears with exponent at most four in each term.
However, most practical optimization problems involve objective functions and constraints that are at most quadratic.
Even when quadratic constraints are embedded into the objective function through penalty or augmented Lagrangian terms\cite{Bertsekas1982}, 
the resulting expressions still contain only up to fourth-order terms, so this restriction is sufficiently practical for real-world applications.

Based on these observations, the choice of transition probability can be summarized as follows.
When the objective function consists only of multilinear terms, the heat bath method can be executed efficiently using the inverse-transform approach, and the optimal-transition Metropolis method can also be applied without difficulty.
Therefore, either the heat bath method or the optimal-transition Metropolis method is appropriate in this case.
For general polynomial objectives, the optimal-transition Metropolis method is preferable for the cases implemented in this study, where each variable appears with exponent at most four.
Otherwise, the conventional Metropolis method is generally the practical choice.

Our results obtained in this study suggest that the proposed approach provides a new and efficient solution method for general integer polynomial optimization problems, which many real-world tasks can naturally be formulated into.
Moreover, the optimal-transition mechanism introduced in this work can be applied without modification to SA for continuous variables, 
and the underlying idea may also be effective for metaheuristics beyond SA.

Finally, we highlight several directions for future work.
Concerning transition probabilities, other efficient update schemes\cite{aoap/1177005652, biomet/83.3.681, PhysRevE.70.056705, PhysRevLett.105.120603, 10.1137/110832288} are also well known.
When the variable range is wide, it is essential for any transition-probability scheme to compute probabilities efficiently.
If these methods are incorporated appropriately into SA, they may further enhance its performance.
Regarding update strategies, our work focused on single-variable updates, where one variable is updated at a time.
Incorporating cluster-based updates, such as those used in the Swendsen–Wang\cite{PhysRevLett.58.86} or Wolff\cite{PhysRevLett.62.361} algorithms,
or allowing simultaneous updates of multiple variables, may offer additional performance improvements.

\section*{Acknowledgment}
A part of this work was performed for Council for Science, Technology and Innovation (CSTI),Cross-ministerial Strategic Innovation Promotion Program (SIP), “Promoting the application of advanced quantum technology platforms to social issues”(Funding agency : QST).

\bibliographystyle{jpsj}
\bibliography{ref}

@ARTICLE{fphy.2014.00005,
author={Lucas, Andrew },
title={Ising formulations of many {NP} problems},
journal={Front. Phys.},
volume={2},
year={2014},
url={https://www.frontiersin.org/journals/physics/articles/10.3389/fphy.2014.00005},
ISSN={2296-424X}
}

@article{Kochenberger2014,
  author    = {Kochenberger, Gary and Hao, Jin-Kao and Glover, Fred and Lewis, Mark and L\"{u}, Zhipeng and Wang, Haibo and Wang, Yang},
  title     = {The unconstrained binary quadratic programming problem: a survey},
  journal   = {Journal of Combinatorial Optimization},
  year      = {2014},
  volume    = {28},
  number    = {1},
  pages     = {58--81},
  month     = {jul},
  issn      = {1573-2886},
  doi       = {10.1007/s10878-014-9734-0},
  url       = {https://doi.org/10.1007/s10878-014-9734-0}
}

@article{Glover2022,
  author    = {Glover, Fred and Kochenberger, Gary and Hennig, Rick and Du, Yu},
  title     = {Quantum bridge analytics I: a tutorial on formulating and using QUBO models},
  journal   = {Annals of Operations Research},
  year      = {2022},
  volume    = {314},
  number    = {1},
  pages     = {141--183},
  month     = {jul},
  issn      = {1572-9338},
  doi       = {10.1007/s10479-022-04634-2},
  url       = {https://doi.org/10.1007/s10479-022-04634-2}
}

@article{SA,
author = {S. Kirkpatrick  and C. D. Gelatt  and M. P. Vecchi },
title = {Optimization by Simulated Annealing},
journal = {Science},
volume = {220},
number = {4598},
pages = {671-680},
year = {1983},
doi = {10.1126/science.220.4598.671},
URL = {https://www.science.org/doi/abs/10.1126/science.220.4598.671},
eprint = {https://www.science.org/doi/pdf/10.1126/science.220.4598.671}
}

@article{SA2,
  author    = {{\v{C}}ern{\'{y}}, V.},
  title     = {Thermodynamical approach to the traveling salesman problem: An efficient simulation algorithm},
  journal   = {Journal of Optimization Theory and Applications},
  year      = {1985},
  volume    = {45},
  number    = {1},
  pages     = {41--51},
  month     = {jan},
  issn      = {1573-2878},
  doi       = {10.1007/BF00940812},
  url       = {https://doi.org/10.1007/BF00940812}
}

@article{Karimi2019,
  author    = {Karimi, Sahar and Ronagh, Pooya},
  title     = {Practical integer-to-binary mapping for quantum annealers},
  journal   = {Quantum Information Processing},
  year      = {2019},
  volume    = {18},
  number    = {4},
  pages     = {94},
  month     = {feb},
  issn      = {1573-1332},
  doi       = {10.1007/s11128-019-2213-x},
  url       = {https://doi.org/10.1007/s11128-019-2213-x}
}

@ARTICLE{9435359,
  author={Tamura, Kensuke and Shirai, Tatsuhiko and Katsura, Hosho and Tanaka, Shu and Togawa, Nozomu},
  journal={IEEE Access}, 
  title={Performance Comparison of Typical Binary-Integer Encodings in an Ising Machine}, 
  year={2021},
  volume={9},
  number={},
  pages={81032-81039},
  doi={10.1109/ACCESS.2021.3081685}}

@article{BOROS2002155,
title = {Pseudo-Boolean optimization},
journal = {Discrete Applied Mathematics},
volume = {123},
number = {1},
pages = {155-225},
year = {2002},
issn = {0166-218X},
doi = {https://doi.org/10.1016/S0166-218X(01)00341-9},
url = {https://www.sciencedirect.com/science/article/pii/S0166218X01003419},
author = {Endre Boros and Peter L. Hammer}
}

@article{Anthony2017,
  author    = {Anthony, Martin and Boros, Endre and Crama, Yves and Gruber, Aritanan},
  title     = {Quadratic reformulations of nonlinear binary optimization problems},
  journal   = {Mathematical Programming},
  year      = {2017},
  volume    = {162},
  number    = {1},
  pages     = {115--144},
  month     = {mar},
  issn      = {1436-4646},
  doi       = {10.1007/s10107-016-1032-4},
  url       = {https://doi.org/10.1007/s10107-016-1032-4}
}

@article{integer_sa_1,
  author        = {Álvaro Rubio-García and Juan José García-Ripoll and Diego Porras},
  title         = {Portfolio optimization with discrete simulated annealing},
  journal       = {arXiv:2210.00807},
  eprint        = {2210.00807},
  archivePrefix = {arXiv},
  primaryClass  = {cond-mat.stat-mech},
  url           = {https://arxiv.org/abs/2210.00807}
}

@article{Abramson1999,
  author    = {Abramson, D. and Randall, M.},
  title     = {A simulated annealing code for general integer linear programs},
  journal   = {Annals of Operations Research},
  year      = {1999},
  volume    = {86},
  number    = {0},
  pages     = {3--21},
  month     = {jan},
  issn      = {1572-9338},
  doi       = {10.1023/A:1018915104438},
  url       = {https://doi.org/10.1023/A:1018915104438}
}

@article{CARDOSO19971349,
title = {A simulated annealing approach to the solution of minlp problems},
journal = {Computers \& Chemical Engineering},
volume = {21},
number = {12},
pages = {1349-1364},
year = {1997},
issn = {0098-1354},
doi = {https://doi.org/10.1016/S0098-1354(97)00015-X},
url = {https://www.sciencedirect.com/science/article/pii/S009813549700015X},
author = {M.F. Cardoso and R.L. Salcedo and S.Feyo {de Azevedo} and D. Barbosa}
}

@article{METROPOLIS,
    title = {{Equation of State Calculations by Fast Computing Machines}},
    journal = {J. Chem. Phys.},
    volume = {21},
    issue = {6},
    pages = {1087-1092},
    year = {1953},
    issn = {1089-7690},
    doi = {https://doi.org/10.1063/1.1699114},
    url = {https://pubs.aip.org/aip/jcp/article-abstract/21/6/1087/202680/Equation-of-State-Calculations-by-Fast-Computing?redirectedFrom=fulltext},
    author = {Nicholas Metropolis and Arianna W. Rosenbluth and Marshall N. Rosenbluth and Augusta H. Teller and Edward Teller}
}

@article{METROPOLIS_HASTING,
    title = {{Monte Carlo sampling methods using Markov chains and their applications}},
    journal = {Biometrika},
    volume = {57},
    issue = {1},
    pages = {97-109},
    year = {1970},
    doi = {https://doi.org/10.1093/biomet/57.1.97},
    url = {https://academic.oup.com/biomet/article-abstract/57/1/97/284580},
    author = {W. K. Hastings}
}

@article{10.1071/PH650119,
    author = {Barker, AA},
    title = {Monte Carlo Calculations of the Radial Distribution Functions for a Proton?Electron Plasma},
    journal = {Australian Journal of Physics},
    volume = {18},
    number = {2},
    pages = {119-134},
    year = {1965},
    month = {04},
    issn = {0004-9506},
    doi = {10.1071/PH650119},
    url = {https://doi.org/10.1071/PH650119},
    eprint = {https://connectsci.au/ph/article-pdf/18/2/119/1347449/ph650119.pdf},
}

@article{ikeuchi2025,
  author        = {Kazuki Ikeuchi and Yoshiki Matsuda and Shu Tanaka},
  title         = {Evaluating the Performance of Direct Higher-Order Formulations in Combinatorial Optimization Problems},
  journal       = {arXiv:2510.24237},
  eprint        = {2510.24237},
  archivePrefix = {arXiv},
  primaryClass  = {cond-mat.stat-mech},
  url           = {https://arxiv.org/abs/2510.24237}
}

@article{PhysRevA.109.032416,
  title = {Annealing for prediction of grand canonical crystal structures: Implementation of $n$-body atomic interactions},
  author = {Couzini\'e, Yannick and Nishiya, Yusuke and Nishi, Hirofumi and Kosugi, Taichi and Nishimori, Hidetoshi and Matsushita, Yu-ichiro},
  journal = {Phys. Rev. A},
  volume = {109},
  issue = {3},
  pages = {032416},
  numpages = {12},
  year = {2024},
  month = {Mar},
  publisher = {American Physical Society},
  doi = {10.1103/PhysRevA.109.032416},
  url = {https://link.aps.org/doi/10.1103/PhysRevA.109.032416}
}

@article{Wang2025,
  author    = {Wang, Bi-Ying and Cui, Xiaopeng and Zeng, Qingguo and Zhan, Yemin and Yung, Man-Hong and Shi, Yu},
  title     = {Speedup of high-order unconstrained binary optimization using quantum ${{\mathbb{Z}}}_{2}$ lattice gauge theory},
  journal   = {Communications Physics},
  year      = {2025},
  volume    = {8},
  number    = {1},
  pages     = {150},
  month     = {apr},
  issn      = {2399-3650},
  doi       = {10.1038/s42005-025-02072-7},
  url       = {https://doi.org/10.1038/s42005-025-02072-7}
}

@article{IJMHEUR.2011.041196,
author = {Glover, Fred and Hao, Jin-Kao and Kochenberger, Gary},
title = {Polynomial unconstrained binary optimisation – Part 1},
journal = {International Journal of Metaheuristics},
volume = {1},
number = {3},
pages = {232-256},
year = {2011},
doi = {10.1504/IJMHEUR.2011.041196},
URL = {https://www.inderscienceonline.com/doi/abs/10.1504/IJMHEUR.2011.041196},
eprint = {https://www.inderscienceonline.com/doi/pdf/10.1504/IJMHEUR.2011.041196}
}

@article{IJMHEUR.2011.044356,
author = {Glover, Fred and Hao, Jin-Kao and Kochenberger, Gary},
title = {Polynomial unconstrained binary optimisation – part 2},
journal = {International Journal of Metaheuristics},
volume = {1},
number = {4},
pages = {317-354},
year = {2011},
doi = {10.1504/IJMHEUR.2011.044356},
URL = { https://www.inderscienceonline.com/doi/abs/10.1504/IJMHEUR.2011.044356a},
eprint = { https://www.inderscienceonline.com/doi/pdf/10.1504/IJMHEUR.2011.044356}
}

@book{Bertsekas1982,
  author    = {Dimitri P. Bertsekas},
  title     = {Constrained Optimization and Lagrange Multiplier Methods},
  publisher = {Academic Press},
  year      = {1982},
doi = {https://doi.org/10.1016/C2013-0-10366-2},
  address   = {New York},
  isbn      = {978-0-12-093480-5}
}

@misc{openjij,
  author  = {Nishimura, Kohji and
             Sakamoto, Yoshiki and
             Shimizu, Taro and
             Suzuki, Kohei and
             Yamashiro, Yu},
  title   = {OpenJij},
  journal = {Zenodo},
  doi     = {10.5281/zenodo.15790495},
 note    = {DOI: 10.5281/zenodo.15790495}
}

@article{PhysRevLett.58.86,
  title = {Nonuniversal critical dynamics in Monte Carlo simulations},
  author = {Swendsen, Robert H. and Wang, Jian-Sheng},
  journal = {Phys. Rev. Lett.},
  volume = {58},
  issue = {2},
  pages = {86--88},
  numpages = {0},
  year = {1987},
  month = {Jan},
  publisher = {American Physical Society},
  doi = {10.1103/PhysRevLett.58.86},
  url = {https://link.aps.org/doi/10.1103/PhysRevLett.58.86}
}

@article{PhysRevLett.62.361,
  title = {Collective Monte Carlo Updating for Spin Systems},
  author = {Wolff, Ulli},
  journal = {Phys. Rev. Lett.},
  volume = {62},
  issue = {4},
  pages = {361--364},
  numpages = {0},
  year = {1989},
  month = {Jan},
  publisher = {American Physical Society},
  doi = {10.1103/PhysRevLett.62.361},
  url = {https://link.aps.org/doi/10.1103/PhysRevLett.62.361}
}

@ARTICLE{4767596,
  author={Geman, Stuart and Geman, Donald},
  journal={IEEE Transactions on Pattern Analysis and Machine Intelligence}, 
  title={Stochastic Relaxation, Gibbs Distributions, and the Bayesian Restoration of Images}, 
  year={1984},
  volume={PAMI-6},
  number={6},
  pages={721-741},
  doi={10.1109/TPAMI.1984.4767596}}

@article{PhysRevLett.105.120603,
  title = {Markov Chain Monte Carlo Method without Detailed Balance},
  author = {Suwa, Hidemaro and Todo, Synge},
  journal = {Phys. Rev. Lett.},
  volume = {105},
  issue = {12},
  pages = {120603},
  numpages = {4},
  year = {2010},
  month = {Sep},
  publisher = {American Physical Society},
  doi = {10.1103/PhysRevLett.105.120603},
  url = {https://link.aps.org/doi/10.1103/PhysRevLett.105.120603}
}

@article{biomet/83.3.681,
    author = {LIU, JUN S.},
    title = {Peskun's theorem and a modified discrete-state Gibbs sampler},
    journal = {Biometrika},
    volume = {83},
    number = {3},
    pages = {681-682},
    year = {1996},
    month = {09},
    issn = {0006-3444},
    doi = {10.1093/biomet/83.3.681},
    url = {https://doi.org/10.1093/biomet/83.3.681},
    eprint = {https://academic.oup.com/biomet/article-pdf/83/3/681/665140/83-3-681.pdf},
}

@article{PhysRevE.70.056705,
  title = {Optimal Monte Carlo updating},
  author = {Pollet, Lode and Rombouts, Stefan M. A. and Van Houcke, Kris and Heyde, Kris},
  journal = {Phys. Rev. E},
  volume = {70},
  issue = {5},
  pages = {056705},
  numpages = {6},
  year = {2004},
  month = {Nov},
  publisher = {American Physical Society},
  doi = {10.1103/PhysRevE.70.056705},
  url = {https://link.aps.org/doi/10.1103/PhysRevE.70.056705}
}

@article{10.1137/110832288,
author = {Chen, Ting-Li and Chen, Wei-Kuo and Hwang, Chii-Ruey and Pai, Hui-Ming},
title = {On the Optimal Transition Matrix for Markov Chain Monte Carlo Sampling},
journal = {SIAM Journal on Control and Optimization},
volume = {50},
number = {5},
pages = {2743-2762},
year = {2012},
doi = {10.1137/110832288},
URL = { https://doi.org/10.1137/110832288},
eprint = { https://doi.org/10.1137/110832288a}
}

@article{aoap/1177005652,
author = {Arnoldo Frigessi and Chii-Ruey Hwang and Laurent Younes},
title = {{Optimal Spectral Structure of Reversible Stochastic Matrices, Monte Carlo Methods and the Simulation of Markov Random Fields}},
volume = {2},
journal = {The Annals of Applied Probability},
number = {3},
publisher = {Institute of Mathematical Statistics},
pages = {610 -- 628},
year = {1992},
doi = {10.1214/aoap/1177005652},
URL = {https://doi.org/10.1214/aoap/1177005652}
}

@article{YaghoutNourani_1998,
doi = {10.1088/0305-4470/31/41/011},
url = {https://doi.org/10.1088/0305-4470/31/41/011},
year = {1998},
month = {oct},
publisher = {},
volume = {31},
number = {41},
pages = {8373},
author = {Yaghout Nourani and Bjarne Andresen},
title = {A comparison of simulated annealing cooling strategies},
journal = {Journal of Physics A: Mathematical and General},
}

@article{Ingber1996,
  author    = {Lester Ingber},
  title     = {Adaptive simulated annealing (ASA): Lessons learned},
  journal   = {Control and Cybernetics},
  volume    = {25},
  number    = {1},
  pages     = {33--54},
  year      = {1996}
}

@article{SZU1987157,
title = {Fast simulated annealing},
journal = {Physics Letters A},
volume = {122},
number = {3},
pages = {157-162},
year = {1987},
issn = {0375-9601},
doi = {https://doi.org/10.1016/0375-9601(87)90796-1},
url = {https://www.sciencedirect.com/science/article/pii/0375960187907961},
author = {Harold Szu and Ralph Hartley}
}

@article{Mahdi2017,
  author    = {Walid Mahdi and Seyyid Ahmed Medjahed and Mohammed Ouali},
  title     = {Performance Analysis of Simulated Annealing Cooling Schedules in the Context of Dense Image Matching},
  journal   = {Computaci{\'o}n y Sistemas},
  volume    = {21},
  number    = {3},
  pages     = {493--501},
  year      = {2017},
  doi       = {10.13053/CyS-21-3-2553}
}

@article{10.1063/1.34823,
    author = {White, Steve R.},
    title = {Concepts of scale in simulated annealing},
    journal = {AIP Conference Proceedings},
    volume = {122},
    number = {1},
    pages = {261-270},
    year = {1984},
    month = {11},
    issn = {0094-243X},
    doi = {10.1063/1.34823},
    url = {https://doi.org/10.1063/1.34823},
    eprint = {https://pubs.aip.org/aip/acp/article-pdf/122/1/261/11806994/261_1_online.pdf},
}

@article{Aarts1985StatisticalC,
  title={Statistical cooling : a general approach to combinatorial optimization problems},
  author={Ehl Emile Aarts and Van Laarhoven},
  journal={Philips Journal of Research},
  year={1985},
  volume={40},
  pages={193-226},
  url={https://api.semanticscholar.org/CorpusID:26843571}
}

@article{S1052623497329683,
author = {Cohn, Harry and Fielding, Mark},
title = {Simulated Annealing: Searching for an Optimal Temperature Schedule},
journal = {SIAM Journal on Optimization},
volume = {9},
number = {3},
pages = {779-802},
year = {1999},
doi = {10.1137/S1052623497329683},
URL = { https://doi.org/10.1137/S1052623497329683},
eprint = { https://doi.org/10.1137/S1052623497329683}
}

@article{Dekkers1991,
  author    = {Dekkers, Anton and Aarts, Emile},
  title     = {Global optimization and simulated annealing},
  journal   = {Mathematical Programming},
  year      = {1991},
  volume    = {50},
  number    = {1},
  pages     = {367--393},
  month     = {mar},
  issn      = {1436-4646},
  doi       = {10.1007/BF01594945},
  url       = {https://doi.org/10.1007/BF01594945}
}

@article{Ali1997,
  author    = {Ali, M. M. and Storey, C.},
  title     = {Aspiration Based Simulated Annealing Algorithm},
  journal   = {Journal of Global Optimization},
  year      = {1997},
  volume    = {11},
  number    = {2},
  pages     = {181--191},
  month     = {sep},
  issn      = {1573-2916},
  doi       = {10.1023/A:1008202703889},
  url       = {https://doi.org/10.1023/A:1008202703889}
}

@article{ALI200287,
title = {A direct search variant of the simulated annealing algorithm for optimization involving continuous variables},
journal = {Computers \& Operations Research},
volume = {29},
number = {1},
pages = {87-102},
year = {2002},
issn = {0305-0548},
doi = {https://doi.org/10.1016/S0305-0548(00)00064-2},
url = {https://www.sciencedirect.com/science/article/pii/S0305054800000642},
author = {M.M. Ali and A. Törn and S. Viitanen}
}

@Inbook{Locatelli2002,
author="Locatelli, Marco",
editor="Pardalos, Panos M.
and Romeijn, H. Edwin",
title="Simulated Annealing Algorithms for Continuous Global Optimization",
bookTitle="Handbook of Global Optimization: Volume 2",
year="2002",
publisher="Springer US",
address="Boston, MA",
pages="179--229",
isbn="978-1-4757-5362-2",
doi="10.1007/978-1-4757-5362-2_6",
url="https://doi.org/10.1007/978-1-4757-5362-2_6"
}

@book{Devroye1986,
  author    = {Devroye, Luc},
  title     = {Non-Uniform Random Variate Generation},
  publisher = {Springer Science \& Business Media},
  year      = {1986},
  isbn      = {978-1-4613-8643-8},
  doi       = {10.1007/978-1-4613-8643-8},
  url       = {https://doi.org/10.1007/978-1-4613-8643-8}
}

@misc{cite_jijmodeling,
  author = {Hiromi, Ishii and Taro, Shimizu and Toshiki, Teramura},
  note   = {presented at The 39th Annual Conference of the Japanese Society for Artificial Intelligence, 2025}
}

@misc{cite_ommx_1,
  author = {Toshiki, Teramura and Taro, Shimizu and Hiromi, Ishii},
  note   = {presented at The 39th Annual Conference of the Japanese Society for Artificial Intelligence, 2025}
}

@misc{cite_ommx_2,
  author = {Toshiki, Teramura and Taro, Shimizu and Hiromi, Ishii},
  title        = {OMMX},
  journal      = {Zenodo},
  note         = {DOI: 10.5281/zenodo.17638431}
}

@article{TIAN1995629,
title = {Nonlinear Integer Programming by Simulated Annealing},
journal = {IFAC Proceedings Volumes},
volume = {28},
number = {10},
pages = {629-633},
year = {1995},
note = {7th IFAC Symposium on Large Scale Systems: Theory and Applications 1995, London, UK, 11-13 July, 1995},
issn = {1474-6670},
doi = {https://doi.org/10.1016/S1474-6670(17)51590-6},
url = {https://www.sciencedirect.com/science/article/pii/S1474667017515906},
author = {Peng Tian and Huanchen Wang and Dongme Zhang}
}

\end{document}